# Strain Induced Adatom Correlations

W.Kappus

wolfgang.kappus@t-online.de

revised v2: 2012-02-07

## Abstract

A Born-Green-Yvon type model for adatom density correlations is combined with a model for adatom interactions mediated by the strain in elastic anisotropic substrates. The resulting nonlinear integral equation is solved numerically for coverages from zero to a limit given by stability constraints. W, Nb, Ta and Au surfaces are taken as examples to show the effects of different elastic anisotropy regions. Results of the calculation are shown by appropriate plots and discussed. A mapping to superstructures is tried. Corresponding adatom configurations from Monte Carlo simulations are shown.

## 1.Introduction

Surface phenomena like nucleation, island growth and ripening have been extensively studied in theory and experiments and are well understood, the details of diffusion, nucleation and ordering on surfaces were reviewed in [1,2]. The correlation of individual adatoms in sub monolayer configurations seems to be less in focus in spite of their potential to act as seeds for later nucleation and growth. We will discuss, if adatom correlations can help to understand structures on surfaces and will analyze under what conditions a 2-dimensional adatom gas can form lattice type structures. One of the methods of Statistical Mechanics to describe correlations in a thermal equilibrium situation is the Born-Green-Yvon (BGY) model [3], applied with extensions in the theory of liquids,e.g. [4]. The BGY model works with all types of interactions but we will show that medium range interactions between particles can create effects beyond next neighbor correlations caused by their attraction. A matter of focus will be the strong influence of coverage on the adatom correlations in the presence of medium range interactions, even in very low coverage situations. As example for medium range adatom interactions we will use substrate strain mediated interactions, knowing that their magnitude - dependent on the substrate-adsorbate combination - can range from small to significant. Elastic interactions between adatoms were reviewed both as relevant amongst many other interactions [5] and of medium range [6]. Recent work emphasized the effect of substrate strain on adatom diffusion [7,8,9,10], growth [11] and surface morphology [12,13]. While much of those results are related to steps and impurities and to flux on the surface we assume in this work a perfect surface with a constant amount of mobile adatoms. There is a span from ab initio calculations of surface parameters used in some of the cited work to the usage of statistical averages within the BGY model and to a continuum elasticity description. We assume for this work - concentrating on medium range effects - that latter is adequate and will try to prove this assumption. If elastic interactions are relevant also their anisotropy will play a role, so we will show the influence of substrate anisotropy to adatom correlations. Metals as substrates are chosen as examples, knowing that they are even less appropriate for a strain only model. The selection of Nb and Ta as substrates is due to their elastic complementary nature, the different adatom correlations predicted on their surfaces can be used to prove the model assumptions. The other substrates chosen represent materials with smaller or greater anisotropy, e.g. Si has an anisotropy similar to that of Ta.

This work reuses and evaluates research of the seventies on adatom density correlations [14] and on elastic interactions of adatoms [15]. The basic idea of the interaction model evaluated is that an adatom exerts a (contracting or expanding) stress field to the surface creating medium ranging strain in the substrate which in turn interacts with the stress field of a neighboring adatom. Like in the case of acoustics a continuum model is used for describing forces and strain, knowing its limitations for the description of small distance effects [16]. For distances of the order two lattice constants the continuum model shows attractive interaction while beyond the interaction is repulsive (except for some crystal directions on substrates with high negative anisotropy). Restricting the substrate to materials with cubic symmetry and the adatom stress to isotropy confines the number of model parameters to a small set: the elastic constants and a stress/temperature ratio. This could help to decide on the applicability of a model focusing on medium range elastic interactions and ignoring other interactions. Calculation of sample adatom configurations with a Monte Carlo approach is used to illustrate the adatom correlations found with the BGY model and (for the case of isotropy) to derive statistical correlations from adatom configurations.

## 2.Basic Assumptions

■ 2.1.Density Distribution by Born-Green-Yvon

Following textbooks of Statistical Mechanics, e.g. [17] the pair distribution $g(\vec{s}_1, \vec{s}_2) \equiv g_{12}$ for adatoms on a surface in Kirkwoods superposition approximation [18] is given by the Born-Green-Yvon integro-differential equation [3]

$$\overrightarrow{\nabla}^{(1)} [\ln g_{12} + u_{12}] = -\theta \int_S g_{23} \, g_{13} \, \overrightarrow{\nabla}^{(1)} u_{13} \, d\vec{s}_3 \qquad (2.1)$$

Here $u_{12} = U_{12}/k_B T$ denotes the adatom-adatom interaction, $\theta$ is the adatom coverage, and $\overrightarrow{\nabla}^{(1)}$ acts on the coordinates of adatom (1) only. The left hand side of (2.1) describes the familiar zero coverage case while the right hand side of (2.1) introduces the nonlinear coverage dependent correlation. Following [14] this equation in the actual 2-d case can be transformed into an integral equation

$$\ln g_{12} + u_{12} = (2\pi)^{-1} \theta \int_S [g_{23} - 1] \, d\vec{s}_3 \int_S g_{43} \left| s_{14} \right|^{-2} \vec{s}_{14} \cdot \overrightarrow{\nabla}^{(4)} u_{43} \, d\vec{s}_4 \quad (2.2)$$

This non linear equation for $g_{12}$ can be solved numerically. The results for the case of $u_{12}$ describing strain induced elastic interactions will be discussed below. A short range interaction will be shortly discussed in addition.

We note the difference between the pair distribution $g(s)$ and the correlation $\nu(s) = g(s) - 1$:

The correlation between two adatoms is expected to decrease rapidly as the distance $s$ increases, while $g(s)$ approaches 1 for large distances.

■ 2.2.Elastic Interactions of Adatoms

Following [15] the elastic energy of a substrate crystal with atoms adsorbed on the surface consists of two parts, first the elastic energy of the distorted substrate and second the energy of adatoms exerting tangential forces on the surface. In a continuum description standard theory of elasticity is used for the substrate and the effects of adatoms are described by a stress field $\pi_{\alpha\beta}(\vec{s}) = P_{\alpha\beta} \, \rho(\vec{s})$ proportional to the local adatom density $\rho(\vec{s})$ and a tensor $P_{\alpha\beta}$ acting only parallel to the surface, so that the elastic energy is

$$H_{el} = \frac{1}{2} \int_V \epsilon_{\alpha\beta}(\vec{r}) \, c_{\alpha\beta\mu\nu} \, \epsilon_{\mu\nu}(\vec{r}) \, d\vec{r} + \int_S \epsilon_{\alpha\beta}(\vec{s}) \, \pi_{\alpha\beta}(\vec{s}) \, d\vec{s} \quad (2.3)$$

The strain field $\epsilon_{\alpha\beta}(\vec{r}) = \frac{1}{2} \left( \nabla_\alpha u_\beta + \nabla_\beta u_\alpha \right)$, related with its displacement field $\vec{u}(\vec{r})$, for a given adatom density $\rho(\vec{s})$ is determined by the requirement of mechanical equilibrium

$$\delta H_{el} / \delta u_\alpha = 0 \quad (2.4)$$

which leads to the starting equations for the displacement field on the surface S and in the bulk V

$$\nabla_\beta \, \sigma_{\alpha\beta} \;=\; \nabla_\beta \, \nabla_\mu \, c_{\alpha\beta\mu\nu} \; u_\nu(\vec{r}) \;=\; 0 \qquad \vec{r} \text{ in } V \qquad (2.5)$$

$$n_\beta \, \sigma_{\alpha\beta} \;=\; n_\beta(\vec{s}) \, \nabla_\mu \, c_{\alpha\beta\mu\nu} \, u_\nu(\vec{s}) \;=\; P_{\alpha\beta} \, \nabla_\beta \, \rho(\vec{s}) \quad \vec{s} \text{ on } S. \qquad (2.6)$$

Introduction of plane wave eigenfunctions (decaying in bulk direction) and solution of eigenvalue $\omega(\vec{\kappa})$ equations for substrate displacement field $\vec{u}(\vec{r})$ finally (for details see [15]) lead to an integral for the interaction of adatoms located at the origin and s using polar coordinates (s,$\phi$) for their distance s = |$\vec{s}$| and pair direction angle $\phi$ with respect to the crystal axes

$$U(s, \phi) \;=\; \sum_p U_p(s, \phi)$$

$$=\; \sum_p (2\pi)^{-1} \, \omega_p \cos(p\phi) \cos\left(p\,\frac{\pi}{2}\right) \int \kappa^2 \exp(-\alpha^2 \, \kappa^2) \, J_p(\kappa s) \, d\kappa \qquad (2.7)$$

where $\omega_p$ denotes a coefficient of a cosine series for eigenvalues $\omega(\kappa)$ and the $\exp(-\alpha^2 \, \kappa^2)$ term implements a cutoff to be discussed in section 2.4. $U(\vec{s}, \vec{s}')$ relates to the elastic energy $H_{el}$ by

$$H_{el} \;=\; \frac{1}{2} \int_S \rho(\vec{s}) \, U(\vec{s}, \vec{s}') \, \rho(\vec{s}') \, d\vec{s} \, d\vec{s}' \qquad (2.8)$$

via the before mentioned local adatom density $\rho(\vec{s})$ .

Equation (2.7) has an exact solution

$$U_p(s, \phi) \;=\; (2\pi)^{-1} \, \omega_p \cos(p\,\phi) \cos\left(p\,\frac{\pi}{2}\right) \Gamma\left(\frac{p+3}{2}\right)$$

$$s^p \, {}_1F_1\left(\frac{p+3}{2}\,;\, p+1\,;\, \frac{-s^2}{4\,\alpha^2}\right) \Big/ \left(2^{p+1}\,\Gamma(p+1)\,\alpha^{p+3}\right) \qquad (2.9)$$

where ${}_1F_1$ denotes the Hypergeometric Function, $\Gamma(p)$ the Gamma function.

Equation (2.7) furthermore has an approximation for distances s large compared to the lattice constant

$$U_p(s, \phi) \;=\; (2\pi)^{-1} \, \omega_p \cos(p\phi) \cos\left(p\,\frac{\pi}{2}\right)(p-1)\,(p+1)\,s^{-3} \;+\; O\left((1/s)^5\right). \qquad (2.10)$$

The dominating isotropic p=0 term of the exact solution (2.9) is negative for small s describing a potential hollow (i.e. an attractive potential), has a positive wall for medium s (i.e. a repulsive potential) and approaches infinity with a $s^{-3}$ law. Intuitively the attraction can be understood as follows: an adatom pushes next neighbors on the substrate aside and the resulting displacement field helps another adatom to share the hollow. The wall indicates the substrate compression (hindering adatoms to jump on the wall) before the substrate can relax at increased distances.

The approximation (2.10) describes a monotonic $s^{-3}$ decrease for all s > 0.

For anisotropic substrates p>0 terms describe the anisotropic part of the interaction and influence the height of the positive wall in dependence of the pair direction angle $\phi$ with respect to the crystal axes. Strongly anisotropic substrates can lead to a weak attractive interaction for certain crystal directions apart from the repulsive wall. Here again the substrate displacements open space for other adatoms.

■ 2.3.Cubic Symmetry

Substrates with cubic symmetry have been analyzed. Such high symmetry keeps the number of variables and the calculation effort low. Furthermore the usage of an isotropic contracting or expanding stress tensor

$$P_{\alpha\beta} \;=\; P\,\delta_{\alpha\beta} \text{ with } (\alpha,\, \beta \in (1,\, 2)) \qquad (2.11)$$

restricts the adatom positions to places with a 4- or 3-fold symmetry and thus to surfaces <001> and <111> . P is a scalar parameter and describes the stress magnitude. $P_{\alpha\beta}$ can be deduced from lattice theory [15].

■ 2.4.Cutoff and Implications

The cutoff length $\alpha$ in equation (2.7) plays a crucial role in defining height and location of the potential wall and the medium range potential. In the following evaluation $\alpha$ is chosen to fit the 2-d Brillouin zone

$$\int \kappa \exp(-\alpha^2 \kappa^2) \, d\kappa = 1 \quad == > \quad \alpha = \frac{\sqrt{2}}{2} \qquad (2.12)$$

■ 2.5.Potential Cap

In the case of a potential hollow at small adatom distances (described in the exact solution (2.9)) the density distribution $g_{12}$ would explode . Therefore a cap $U_w = U(s_w)$, the wall height, is introduced for distances s less than the wall maximum $s_w$ location replacing the exact $U(s)$ . See Fig. 1. For the anisotropic case an adapted $U(s, \phi) = U_w + U_{wp} \cos(p\phi) \frac{s}{s_w}$ cap is used. Such cap of course ignores the attractive short range potential region and is an artificial means to enable an equilibrium model describing mid range effects while ignoring short range effects. An equilibrium model appears valid as long as nucleation is not existent or just a small perturbation of the equilibrium state.

■ 2.6.Wall Height and Nucleation

The wall height $U_w$ depends on $\omega_p$ which in turn depends on the elastic constants and the stress field parameter P, the value of which is unknown. Since the pair distribution function $g_{12}$ depends (beside the coverage $\theta$) on the temperature scaled interaction $u_{12} = U_{12}/k_B T$ the scaled wall height $u_w = U_w/k_B T$ can be chosen as a model parameter.

A high wall $u_w \gg 1$ would keep adatoms apart and thus prevent nucleation while a low wall $u_w \approx 1$ would support nucleation. To simulate (the non-equilibrium phenomenon) nucleation in the present equilibrium description we allow for a non vanishing g(s) in the $s < s_w$ region by choosing the (artificial) value $u_w = 5$ in the calculations below. On anisotropic substrates $u_w$ will depend on $\phi$ , so 5 is chosen as average wall height.

■ 2.7.Notation

The combination of different disciplines urged some notation changes compared with [14,15]. Furthermore some notations are introduced or explained:

- while $\rho$ is used for adatom density modes within the elastic interaction section, $\theta$ will denote the coverage ($\rho$ within [14]). $\theta_0$ and $\theta_c$ denote the limiting coverages for the 2 different elastic interaction variants under study

- $\vec{\kappa}$ is the wave vector

- for the adatom pair interaction $U(\vec{s},\vec{s}')$ will be used instead of $W(\vec{s},\vec{s}')$ and $u_{ij} \equiv U(\vec{s}_i ,\vec{s}_j)/k_B T$. $\vec{u}$ denotes the substrate displacement field

- the (hypergeometric) interaction according to the exact case (2.9) will be denoted $u_1$, this case together with the cap according to section 2.4 will be denoted $u_2$, the interaction according to the long distance case (2.10) will be denoted $u_3$. An additional short range interaction $u_4$ will be introduced in section 3.6

- $u_0$ denotes the coefficient in the long distance case $u_3 = u_0 \, s^{-3}$

- the maximum location of interaction $u_1$ will be denoted $s_w$

- $s_0$ is defined by $u_3(s_0) = 1$

- $s_h$ is defined by $u_2(s_h) = 1$. See Fig. 1 for illustration

- elastic constants $c_{\alpha\beta\mu\nu}$ for the cubic system read in Voigts notation $c_{11}$, $c_{12}$, $c_{44}$. For convenience we use the anisotropy parameter $\zeta = (c_{11}-c_{12}-2c_{44})/c_{44}$

- $P_{\alpha\beta}$ denotes the surface stress tensor, P stress parameter within $P_{\alpha\beta} = P\delta_{\alpha\beta}$ while the small p labels the coefficients $\omega_p$ of the $\omega(\phi)$ cosine series expansion

- $\alpha$ is used as cutoff length in (2.12) and also as index of elastic constants $c_{\alpha\beta\mu\nu}$

- <100> stands for one of the equivalent directions <010>, <0-10>, <-100> of the cubic <001> surface

# 3.Calculation

- ### 3.1.Evaluating the Elastic Eigenvalue Equations

Details of finding proper $\omega_p$ have been given in [15]. The eigenvalue equations have been reprogrammed and the previous results ($\omega_0$, $\omega_4$, $\omega_8$, $\omega_6$, $\omega_{12}$) were confirmed apart from current values of the elastic constants [19] and improved numerical accuracy. For isotropic substrates $\omega_0$ has the closed solution

$$\omega_0 = -c_{11} P^2 \Big/ (2 c_{44} (c_{11} - c_{44})) \quad (3.1)$$

where the strong role of the stress parameter P is apparent.

Due to scaling the numerical $\omega_p$ results for the general anisotropic cases will be given in $P^2/c_{44}$ units.

A few substrates have been selected for this study, representing different ranges of anisotropy: W as isotropic material to demonstrate validity of the algorithms used , Nb with moderate positive anisotropy , Ta (like Si and Pt) with moderate negative anisotropy, Au (like Ag and Cu) with strongly negative anisotropy. Tab. 1 shows the resulting $\omega_p$.

```
Substrate   c₁₁    c₁₂    c₄₄     ξ        ω₀       ω₄       ω₈
    Au      191.   162.   42.2   -1.313   -0.895   -0.094   -0.0013
    Nb      245.   132.   28.4    1.979   -0.455    0.0205   0.0009
    Ta      264.   158.   82.6   -0.7167  -0.790   -0.0336  -0.0006
    W       523.   203.   160.    0.      -0.720    0.       0.
```

Table 1. Substrate Elastic Constants $c_{ik}$ (GPa) from [19], anisotropy $\zeta=(c_{11}-c_{12}-2c_{44})/c_{44}$ and coefficients $\omega_p$ (in $P^2/c_{44}$ units) of the cosine series expansion of $\omega(\phi)$ on <001> surfaces.

We note the restriction $\zeta>-2$ for the eigenvalue equations.

- ### 3.2.Evaluating the Pair Distribution Function

Computing $g(\vec{s}_1,\vec{s}_2)=g_{12}$ for different $\theta$ from (2.2) is straightworward starting iterations from zero coverage and increasing $\theta$ slowly. The iteration step leading from $g_n$ to $g_{n+1}$ for coverage $\theta_i$ is

$$g_{n+1} = \lambda \exp(-u + \theta_i F(g_n)) + (1-\lambda) g_n, \quad (3.2)$$

where $\lambda$ denotes a damping parameter. $\lambda$ can be chosen 1 for small coverages $\theta_i$ but must be reduced towards 0.1 when $\theta_i$ approaches the critical coverage $\theta_c$. Limited computing resources implied a coarse grid for $s_{12}$ ($\sim 10^3$ grid points), so the numerical accuracy for $g_{12}$ is in the low % area. Convergence of (2.2) is good for small $\theta$ but strongly decreasing when approaching $\theta_0$ in the isotropic case and the limiting coverage $\theta_c$ , a fraction of $\theta_0$ , in anisotropic cases. As a rule of thumb $\theta_c$ increases linearly between $\zeta=-2$ and $\zeta=0$ and decreases with a $(\zeta+2)^{-1}$ power law for $\zeta>0$ on <001> surfaces. Appropriate damping is necessary beyond the small $\theta$ region. Divergence of (2.2) means explosive growth of g(s) peaks indicating the limit of the equilibrium theory used. A Computer Algebra System has been used to perform the numeric calculations and the graphics presentation.

- ### 3.3.Isotropic Reference

In the case of an isotropic substrate we have

$$u(s) = u(|\vec{s}|), \quad g(s) = g(|\vec{s}|). \quad (3.3)$$

The pair distribution equation (2.2) in this case reads [14]

$$\ln g(R) + u(R) =$$

$$\theta \int_0^\infty r[g(r)-1] \, dr \int_0^\infty g(l) u'(l) \, dl \int_0^{2\pi} \Theta\Big[1 - \sqrt{R^2 + r^2 - 2Rr\cos\beta}\Big] \, d\beta, \quad (3.4)$$

where $\Theta$ denotes the Heaviside Theta function.

Unfortunately [14] contains printing errors as detected during this work: the ' (differentiation symbol) was omitted in eq. (8) and (9) and a $\pi$ was omitted at the upper integral limit in eq. (8). Also the second part of eq. (9) was not verified and not used in the current calculations.

■ 3.4.Pair Distribution Function of an Isotropic Short Range Interaction

Before going into the details of the medium term elastic interaction we need to verify that the pair distribution function caused by short term interactions is well separated from the pair distribution caused by medium term interactions.For this purpose we take as an example interaction

$$u_4(s) = s^{-12} - 2\,s^{-6}. \quad (3.5)$$

chosen for its simplicity without deeper physical background. $u_4(s)$ is strongly repulsive at small distances, has an attractive minimum $u_4(1)=-1$ and approaches zero rapidly at distances $s<s_w$ (see Fig. 1). The coverage dependent pair distribution function $g(s,\theta)$ for interaction $u_4(s)$ calculated with (3.4) is shown in Fig.2.a. It shows its first maximum at lattice constant distance $s=1$, a minimum emerging with increasing coverage at $s\approx1.8$ and a second maximum emerging at $s\approx2.2$. Keeping in mind the limits of the BGY model for large coverages this can be interpreted by a distorted hexagonal package of adatoms neighbors. We note that in the low coverage regime $\theta<0.05$ discussed below $g(s,\theta)\approx g(s,0)$ and conclude sufficient separation between the short- and medium term regimes discussed.

■ 3.5.Pair Distribution Function of an Isotropic Elastic Interaction

We note from (3.4) that the pair distribution function $g(s)$ with a medium term isotropic interaction

$$u_3(s) = u_0\,s^{-3} \quad (3.6)$$

scales, i.e.

$$g(s,\ u_0,\ \theta) = g\left(\tau s,\ \tau^3\,u_0,\ \tau^{-2/3}\,\theta\right) \quad (3.7)$$

with a scaling factor $\tau$. So just one evaluation of (3.4) with (3.6) covers all values of the interaction constant $u_0$.
The relations

$$s_0 = u_0^{1/3} \ \text{and} \ \theta_0 = u_0^{-2/3} \quad (3.8)$$

help to find appropriate step widths and coverage ranges when solving (2.2) and (3.4) numerically and to discuss results. The exact interaction $u_1(s)$ lacks this scaling feature.

Evaluation of the isotropic $g(s)$ was done by solving (3.4) numerically. Reduction in dimension compared to the general anisotropic case leads to much faster numeric results. Furthermore the 2 different algorithms have served as mutual test cases during programming.

The effects of both the approximate $u_3$ and the capped exact interaction $u_2$ on $g(s)$ are shown in Figs. 2.b and 2.c for $u_0 \approx 51$ (a value leading to the above mentioned wall height of $u_w=5$) and stepwise increasing coverage $\theta$.

In both cases a main peak emerges with increasing $\theta$. In the $u_3$ case the peak location tends to $s_0 = u_0^{-3}$ when $\theta$ approaches $\theta_0=u_0^{-2/3}$ while in the $u_2$ case the first peak location $s_h$ is shifted towards a larger adatom distance defined by $u_2(s_h)=1$. The height of the main peak stays below 1.5 which turns out to be a stability limit within the numerical calculations.

Physically the nearest adatoms are pressed by the increasing coverage $\theta$ to positions where the interaction $U$ equals $k_B T$.

Coverages $\theta > \theta_0$ are unstable in this equilibrium theory.

$g(s)$ shows a damped oscillating behavior, further (smaller) peaks emerge with increasing $\theta$ at $N*s_0$ distances indicating the location of next nearest neighbors.

Fig. 2.c shows a strong increase of $g(s)$ in the cap region for increasing $\theta$ indicating that some adatoms have climbed over the wall $u_w$ , i.e. that nucleation took place . The details of nucleation effects, however, are beyond the current elastic model since short range interactions (between adatoms in the hollow) are not accounted for.

■ 3.6.Presentation of Results for Cubic <001> Surfaces

For every substrate material evaluated the presentation contains 360° contour plots with the reference adatom in the origin and value legends in the lower left quadrant. The contour plots range from s = 0 to 22 to emphasize the relevant region. Furthermore g(s, $\theta$) evolution plots are shown. Especially:
- an u(s) contour plot, dark colors represent high values, light colors represent low (sometimes negative) values,contours changing in 1- steps
- a $g_0$(s) contour plot representing the coverage $\theta$ = 0 starting pair distribution, dark colors represent high values, light colors represent low values, contours changing in 0.1 steps
- a g(s) contour plot for the limiting coverage $\theta_c$ , dark colors represent high values, light colors represent low values, contours changing in 0.1 steps
- a semi-logarithmic $g_{<100>}$(s,$\theta$) plot for s ∥ <100>, showing the evolution of g(s) with increasing coverage $\theta$
- a semi-logarithmic $g_{<110>}$(s,$\theta$) plot for s ∥ <110>, showing the evolution of g(s) with increasing coverage $\theta$

Figs. 3 to 6 show the resulting u(s) and g(s) for W, Nb, Ta, Au substrates respectively. As already mentioned the substrates were chosen to represent regions of elastic anisotropy and the results should give an idea how the adatom pair distribution on substrates with similar elastic properties would look like.

■ 3.7.Presentation of Results for Cubic <111> Surfaces

<111> surfaces turn out to be much closer to isotropy than <001> surfaces. Substrates like Nb with a smaller anisotropy show results on <111> very similar to the isotropic W. Therefore we restrict the presentation to the most anisotropic case, Au. The coefficients of the cosine series expansion $\omega(\phi)$ for Gold are $\omega_0$=-0.845, $\omega_6$=-0.0082, $\omega_{12}$=0.000317. The figure sequence presented is like in section 3.6. Different are only the crystal directions, now <1-10> and <-1-12>.

Figs. 7 shows the resulting u(s) and g(s) for Au.

■ 3.8.Complementing Monte Carlo Simulations

The pair distribution function g(s) can be interpreted as statistical average over many adatom position samples. Monte Carlo simulations of adatoms interacting with the capped $u_2$ have been performed. In accordance with the continuum model used for describing the interaction, a grid-less algorithm has been used. Periodic boundary conditions were applied. The area size of 60 units was chosen to keep the computing time in the range of hours while the interaction u(s=60) has decreased well below 0.001. Starting from a random k member adatom configuration {$\vec{s}_{i,0}$}, step n + 1 {$\vec{s}_{i,n+1}$} evolves from step n {$\vec{s}_{i,n}$} by

$$\vec{s}_{i,n+1} = \vec{s}_{i,n} + \chi \sum_{j=1}^{k} \vec{\nabla}^{(j)} u(\vec{s}_{i,n}, \vec{s}_{j,n}), \quad (3.9)$$

with an appropriate parameter $\chi$. So an adatom moves around under the forces of all its neighbors until all forces are balanced. Convergence is achieved up to the critical coverages $\theta_c$. Beyond $\theta_c$ the differences $\vec{s}_{i,n+1}$-$\vec{s}_{i,n}$ increase after a certain minimum and adatom movement and nucleation continues. The method chosen does not account for thermal movement of the adatoms and therefore will provide only an approximation to their thermal equilibrium. Figs. 3.d and 4 to 7.f show adatom position samples from such Monte Carlo simulations for the materials evaluated and coverages $\theta_c$. Paired or clustered objects with a distance < $s_w$ (representing nucleated adatoms) are depicted in red, others in blue color.

Though the statistical average over many adatom configurations in general requires computing resources unavailable, the special case of elastically isotropic W allows such calculation. Fig. 3.e shows the result.

# 4.Discussion

■ 4.1.Tungsten

W as an isotropic substrate has been chosen to verify the consistence between the dedicated isotropic algorithm and the general anisotropic algorithm. In addition to Fig. 2.c (which shows g(s,$\theta$) for W) the g(s,$\phi$) contour plot for the critical coverage $\theta_c$=0.05 is shown in Fig. 3.c Oscillations around g(s)=1 with a wavelength around $s_h$ are now obvious also for large distances s since the value 1 marks the boundary between g(s)<1 and g(s)>1. A distorted hexagonal super-lattice of adatoms with site distances around s≈5 (avoiding the s≈7.5 distance, the 2nd next neighbor distance in a perfect hexagonal lattice) would map to these results . We note $\theta$=0.046 for a perfect 5-site hexagonal lattice. A value of g(2.5,$\theta_c$)≈0.075 indicates the onset of nucleation. Fig. 3.d shows a corresponding adatom sample. About 3/10 of the adatoms have paired/clustered (distance < $s_w$) at $\theta_c$(W). Fig. 3.e shows the corresponding pair correlation averaged over 40 samples. The peak distances at s≈5 and s≈10 correspond to those in Fig. 2.c while the peak heights reflect the absence of thermal movement in the case of the Monte Carlo simulation. Nucleation effects are not visible in Fig. 3.e since positions with s<1 have been eliminated.

■ 4.1.Niobium <001>

Niobium is a substrate with moderate positive anisotropy. The u(s) contour in Fig. 4.a shows a rounded square with wall maxima in the <100> direction. The wall heights are $u_{w<100>}$≈5.57 and $u_{w<110>}$≈4.43. The associated $g_0$(s) contour in Fig. 4.b thus shows higher values in the <110> direction. The g(s) contour for $\theta_c$ in Fig. 4.c shows its maximum in <110> direction at s≈5 (beyond $s_h$≈ 4) with an adjacent ridge. Towards <100> a flat saddle shows up at s≈6. Drops in the <210> direction indicate weak g(s) oscillations around the limiting value 1. The small deviations from the 90° rotation symmetry are due to the limited accuracy of the contour routine. $g_{<100>}$(s,$\theta$) in Fig. 4.d shows a minimum at s≈10 not visible in the contour plot. $g_{<110>}$(s,$\theta$) in Fig. 4.e shows beside the maximum emerging towards $s_h$ a secondary one emerging towards s≈10 sites. In combination the maxima at s≈5 and 10 in <110> direction and at s≈6 in <100> direction indicate the formation of a square superstructure like (<440>,<4-40>). The lines represent coverages $\theta$ from 0 to 0.025 in 0.005 steps. Values of $g_{<100>}$(2.5,$\theta_c$)≈0.013 and $g_{<110>}$(2.5,$\theta_c$)≈0.044 indicate the onset of nucleation from the <110> direction. Fig. 4.f shows a corresponding adatom sample. About 1/8 of the adatoms have paired/clustered (distance < $s_w$) at $\theta_c$(Nb). Adatoms form rather <110> directed chains than a square lattice. Such chains are also compatible with the g(s) contour.

■ 4.3.Tantalum <001>

Tantalum is a substrate with moderate negative anisotropy. The wall heights are $u_{w<100>}$≈4.45 and $u_{w<110>}$≈5.55. Both the u(s) and the $g_0$(s) contours in Figs. 5.a and 5.b look like 45° rotated versions of Nb. This comes from the approximately opposite value of $\omega_4/\omega_0$ , see Tab. 1 . The g(s) contour of Ta in Fig. 5.c looks not quite like a 45° rotated version of Nb, there is a slight minimum in the <100> direction at s≈15 and the g<1 area in the <110> direction extends further .Those differences are addressed to the respective $\omega_8$ values (no sign change). $g_{<100>}$(s,$\theta$) in Fig. 5.d shows the maximum emerging towards s≈5 (beyond $s_h$≈4). $g_{<110>}$(s,$\theta$) in Fig. 5.e shows a small maximum emerging towards s≈7 and a minimum emerging towards s≈10. In combination the maxima at 5 in <100> direction and at 7≈5*$\sqrt{2}$ in <110> direction indicate the formation of a square superstructure like (<500>,<050>). We note repulsions $u_{<100>}$(5)≈0.3 and $u_{<110>}$(7)≈0.26 in favor for such structure. The lines represent coverages $\theta$ from 0 to 0.025 in 0.005 steps. Values of $g_{<100>}$(2.5,$\theta_c$)≈0.041 and $g_{<110>}$(2.5,$\theta_c$)≈0.013 indicate the onset of nucleation from the <100> direction. Fig. 5.f shows a corresponding adatom sample. About 1/10 of the adatoms have paired/clustered (distance < $s_w$) at $\theta_c$(Ta). Adatoms seem to prefer <100> directed chains over a square lattice. Such chains are also compatible with the g(s) contour.

■ 4.4.Gold <001>

Gold is a substrate with high negative anisotropy. The u(s) contour in Fig. 6.a shows wall maxima in the <110> direction and an attractive interaction in the <100> direction from about 5 sites apart from the origin adatom. The wall heights are $u_{w<100>}$≈3.63 and $u_{w<110>}$≈6.37. Potential minimum is about 6 sites apart. Accordingly $g_0$(s) in Fig. 6.b shows values >1 in the <100> direction. The g(s) contour in Fig. 6.c shows a maximum in <100> direction at $s_h \approx 5.5$, adjacent ear like mountains directed towards <310> and <3-10>, a flat saddle and a long g<1 tail towards <110>. Fig. 6.d shows the growing maximum of $g_{<100>}(s,\theta)$ and Fig. 6.e an emerging turning point of $g_{<110>}(s,\theta)$. The lines represent coverages $\theta$ from 0 to 0.016 in 0.004 steps, the critical $\theta_c$ is far below the expected $s_h^{-2}$ value of appr. 0.033 for a square lattice. There is no evidence for a square type super-lattice but <600> chains and <620><6-20> zig-zag chains or forks could map to g(s). Reason for such preference of chains over square patterns could be the strong diagonal repulsion u(<660>)≈0.22 compared to u(<620>)≈0.02 and u(<600>)≈-0.05. Values of $g_{<100>}(2.5,\theta_c) \approx 0.036$ and $g_{<110>}(2.5,\theta_c) \approx 0.002$ indicate the onset of nucleation almost from the <100> direction. Solving (2.2) for Au shows a weak stability towards higher coverages associated with an increasing region of g(s) significantly deviating from 1, i.e. a depletion of adatoms in the <110> direction and a surplus in the <100> direction (increasing the necessary grid size for the calculation and requiring slow changes of $\theta_i$ in (3.2)). Fig. 6.f shows a corresponding adatom sample at $\theta_c(Au<001>)$. About 1/25 of the adatoms have paired/clustered. Adatoms form short, forked <100> directed chains and voids indicate the rather low critical coverage $\theta_c$. The typical distance of adatoms in a chain is about 4 compared with 5.8, the peak location in Fig. 6.d. This difference is explained by the neglection of thermal movement in the Monte Carlo calculation (3.9) noting a still low nearest neighbor repulsion of $u_{<100>}(4) \approx 0.3$.

- **4.5. Gold <111>**

Compared with the <001> surface the <111> surface of Au is elastically much smoother. The u(s) contour in Fig. 7.a shows a nearly isotropic wall with maxima in the <-1-12> directions. The wall heights are $u_{w<1-10>} \approx 4.95$ and $u_{w<-1-12>} \approx 5.05$. Accordingly $g_0(s)$ values in Fig. 7.b are slightly higher in the <1-10> direction. The g(s) contour in Fig. 7.c shows a maximum in <1-10> direction at $s \approx 4.8$ (>$s_h \approx 4.2$), a secondary maximum at $s \approx 10.5$ within a honeycomb-like g>1 ring. In <-1-12> direction a small maximum at $s \approx 5$ is followed by a minimum at $s \approx 7.5$. Fig. 7.d shows the growing maximum of $g_{<1-10>}(s,\theta)$ and Fig. 7.e a smaller maximum of $g_{<-1-12>}(s,\theta)$ at $s \approx 5$ and a minimum emerging at $s \approx 7.5$. The lines represent coverages $\theta$ from 0 to $\theta_c=0.04$ in 0.008 steps. The critical coverage $\theta_c$ is much higher than in the <001> surface case. The g(s) contour in Fig. 7c could tried to be mapped to an aligned hexagonal super-lattice of adatoms with distances of 5 sites. We note that the coverage $\theta_c=0.04$ approaches the value of 0.046 for a full 5-site super-lattice coverage. The minima at $s \approx 7.5$, however, indicate a tendency towards chains competing against a hexagonal lattice. The onset of nucleation is indicated by a value of $g(2.5,\theta_c) \approx 0.055$. Fig. 7.f shows a corresponding adatom sample at $\theta_c(Au<111>)$. About 1/4 of the adatoms have paired/clustered. Adatoms form a <-1-12> aligned hexagonal lattice with voids.

- **4.6. General Aspects**

Lots of assumptions and approximations have been used for this model
- omitting all interactions but the elastic ones
- continuum theory for the substrates instead of a lattice theory, known to be inadequate for describing short range effects
- a cosine series expansion of the $\omega(\kappa)$ eigenvalue with only 3 members
- an isotropic cutoff length with far reaching implications on the interaction characteristics [16]
- the superposition approximation and the BGY theory, know for its limits in the theory of liquids [4]
- the lack of knowledge on the size of the stress parameter P
- the height of the $u_w$ barrier
- the grid size used in numerical calculations
while some model conclusions (e.g. distances) appear quite strong. The validity of those approximations is limited but other interactions as discussed in [5] could be used to calculate an improved pair distribution function.

The map between the pair distribution g(s) and adatom structures remains vague, the limited confidence in the validity of especially the superposition approximation gives some more credibility to the Monte Carlo configuration samples. On the other side the Monte Carlo method would require some hundred samples to construct a meaningful pair distribution in the general anisotropic case, so both methods have pros and cons. Inclusion of thermal effects in the Monte Carlo method would even increase the computing resources required. An on-grid Monte Carlo method could improve the comparability with experiments at the expense of consistency with the continuum theory used for calculating the pair distribution.

The onset of nucleation when the coverage approaches $\theta_c$ needs commenting: we recall that at $\theta_c$ the first peak location is near $s_h$, defined by $u_2(s_h) = 1$. So the potential of an adatom surrounded by 4 or 6 neighbors may reach or exceed the wall height $u_w=5$ chosen for the calculations. Jumping over the barrier (and staying within the $s_w$ circle) is then much more probable than in a low coverage case. The difference in nucleation figures between the (2.2) model and the Monte Carlo simulation is explained by noting that the first is a thermal equilibrium approach while the latter is a mechanical equilibrium approach. Taking into account the artificial potential cap used the (2.2) model provides an equilibrium aspect of the non-equilibrium nucleation mechanisms.

Since for given adatom/bulk combinations U(s) is a function of the stress parameter P, the temperature via $u(s)=U(s)/k_BT$ strongly influences the wall height $u_w \sim T^{-1}$ and also adatom distances like $s_0 \sim u_0^{1/3} \sim T^{-1/3}$. Accordingly the critical coverage $\theta_0 \sim T^{2/3}$ which, with some caution, would also hold for $\theta_c$. Since all lengths discussed depend on the basic assumption of $u_w=5$ they refer to a (yet unknown) temperature T. Au as a most noble metal may have a small stress parameter P, which would require low temperatures to see the effects discussed. Ag or Cu with similar anisotropies but possibly higher P could offer effects at complementary temperature ranges.

# 5.Speculation

Anyhow it is attractive to discuss the consequences if the model assumptions (or parts of them) and their evaluation could be verified for certain adatom-substrate systems:
- adatoms would form quasi-particles with their associated strain field via their stress parameter P. A certain magnitude of P would be necessary to create significant effects
- diffusion of adatom quasi-particles would differ from diffusion of bare adatoms by their differences in activation energy, adatoms would prefer jumps within their hollow
- formation of adatom superstructures would depend on the elastic constants and on the coverage
- formation of the single adatom density distribution and nucleation would compete. Nucleation would be hampered by the repulsive barrier around each adatom but promoted with increasing coverage
- adatom dimers (N-mers) could form bounded states within the elastic hollow even in the absence of other interactions
- the formation of such adatom clusters would depend on the crystal direction on anisotropic substrates, also their shape may reflect this
- adatom N-mer clusters could create their individual strain field either by a simple superposition of the member fields or by additional mechanisms like lattice mismatch. Such clusters could exert stronger stress (hindering further nucleation) or even opposite sign stress (promoting further nucleation). Cluster distances would be a measure for their average (generally anisotropic) stress field strength
- adatom superstructures could grow into cluster superstructures via nucleation at adatom seeds and subsequent ripening
- vacancies, steps, kinks would influence the adatom correlations
- small changes of temperature could change adatom structures .

# 6.Summary

Using a statistical model for the correlation of adatoms subject to substrate mediated strain the conditions are evaluated under which adatom assemblies can form regular structures. In the course of the calculations some assumptions and simplifications have been made and are rated as follows:

- the Born-Green-Yvon model seems appropriate to describe the pair distribution function in the low coverage regime analyzed, the Monte Carlo simulations provided support the results apart from thermal movement

- support for the effects of substrate strain are given in the references

- the relevance of short range attractions was not put in question. Substrate strain - under certain conditions - may create additional medium range effects on the adatom pair distribution

- the elastic continuum model used seems appropriate for the medium range effects analyzed. Short range effects, shown to be separated from the medium range, would require a lattice approach. An ab initio lattice approach could also deliver the $P_{\alpha\beta}$ surface stress tensor

- the capped potential assumption seems to be a valid method to maintain an equilibrium description for a metastable problem. It, however, does not allow to treat short range interactions simultaneously

- the unknown magnitude of the stress parameter P limits the value of the model. As P determines the length scale and the temperature range of effects this is a serious defect. Restriction to high symmetry adatoms places seems more appropriate

- the choice of the wall height $u_w=5$ seems appropriate in demonstrating the balance between adatom structures and nucleation

- consequences of the model assumptions are formulated as speculation since insufficiently founded. E.g. the statement "adatoms would prefer jumps within their hollow" is supported by [9] but the high activation barrier for dimer dissociation can also be a consequence of the short range attraction of adatoms.

In summary this paper follows a chain of arguments: elastic interactions have medium range, medium range interactions can lead to relevant correlation effects, correlation effects can influence nucleation and growth.

The proof of existence of the described adatom structures and of possible influences to subsequent nucleation and growth is beyond this work.

## Acknowledgement

This work is dedicated to the memory of my mother, gone during evaluation.

Many thanks to Prof. F.Wegner for mentorship and encouragement. The author is indebted also to the unknown referees for their guidance and references proposed.

Many thanks also to my family for patience and support.

# **Appendix**

■ **Figures**

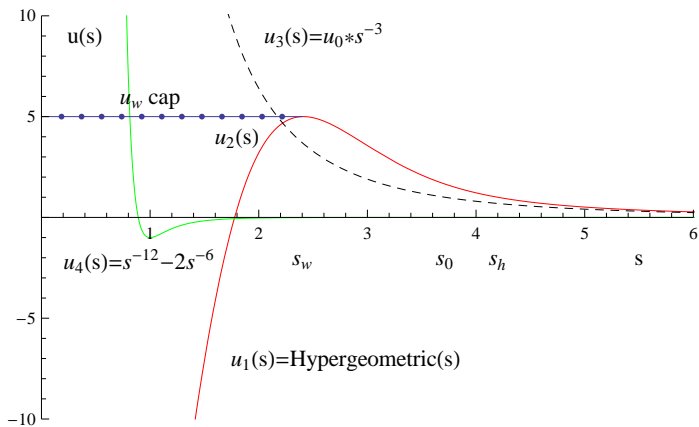

Fig.1 Illustration of interaction types $u_x$ and lengths $s_x$

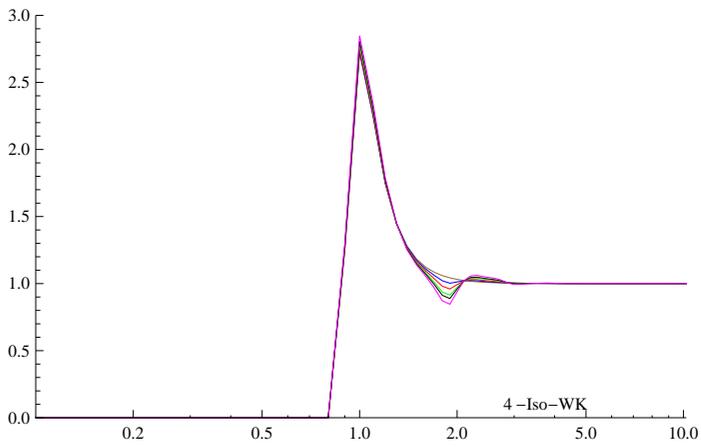

Fig. 2.a Pair distribution function g(s,θ) with the short range interaction $u_4$(s) and coverages $θ$ between 0 and $θ_c$=1.0

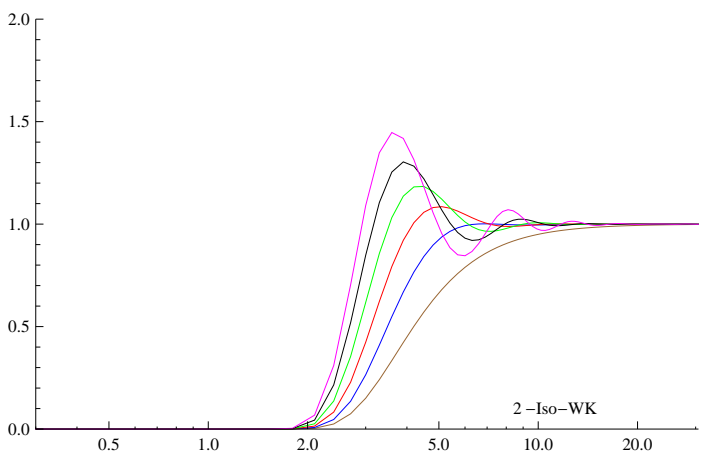

Fig. 2.b Pair distribution function g(s,θ) with the isotropic approximate interaction $u_3$(s) and coverages $θ$ between 0 and $θ_0$=0.075

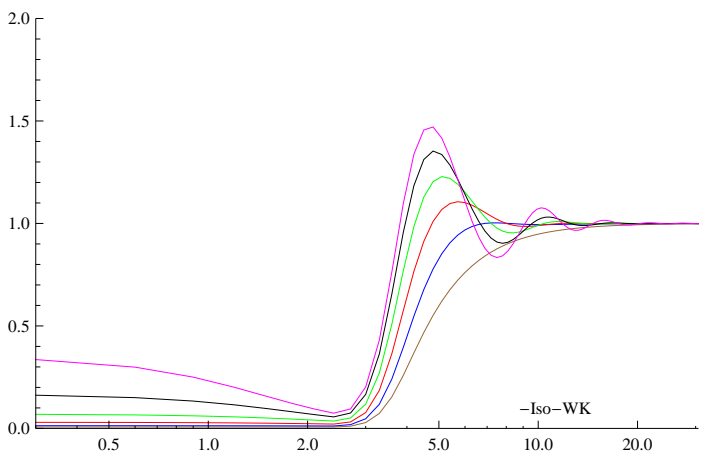

Fig. 2.c Pair distribution function g(s,θ) with the capped exact interaction $u_2$(s) and coverages $θ$ between 0 and $θ_c$=0.05

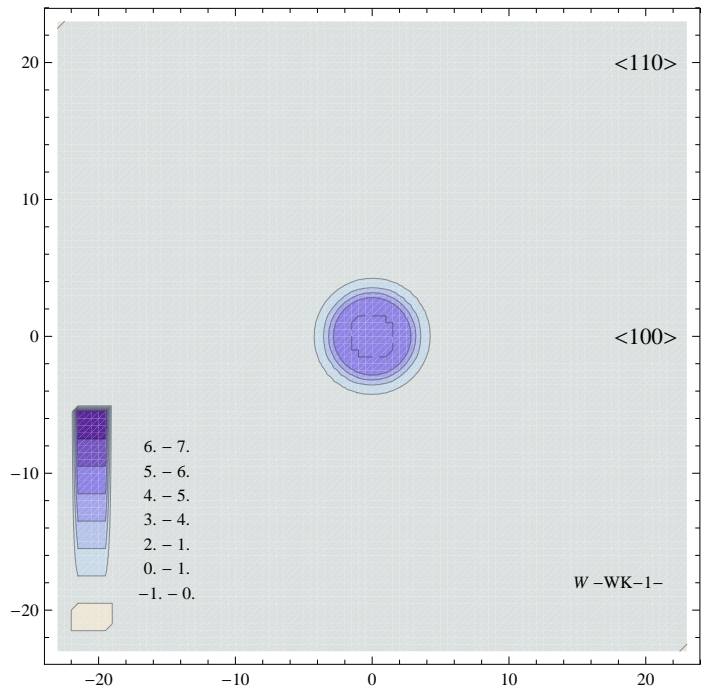

Fig. 3.a u(s) contour plot for W

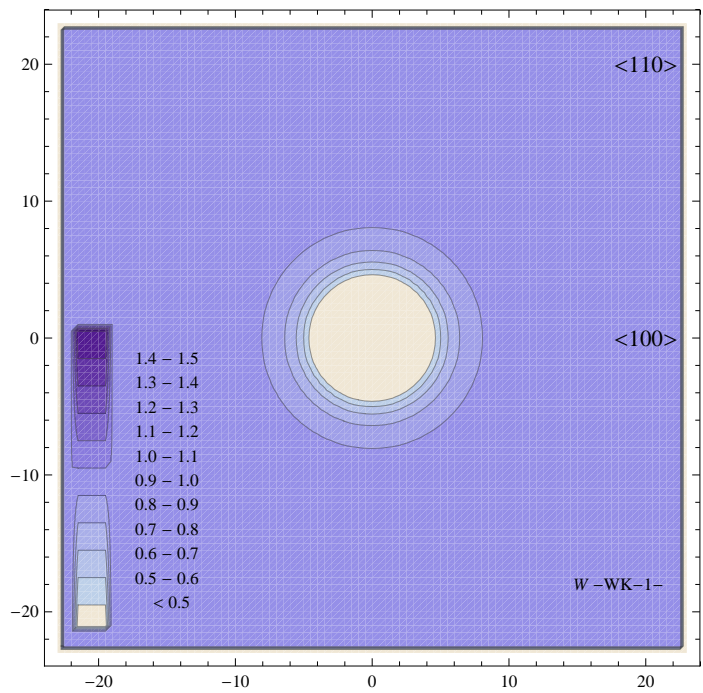

Fig. 3.b Zero coverage $g_0(s,0)$ contour plot for W

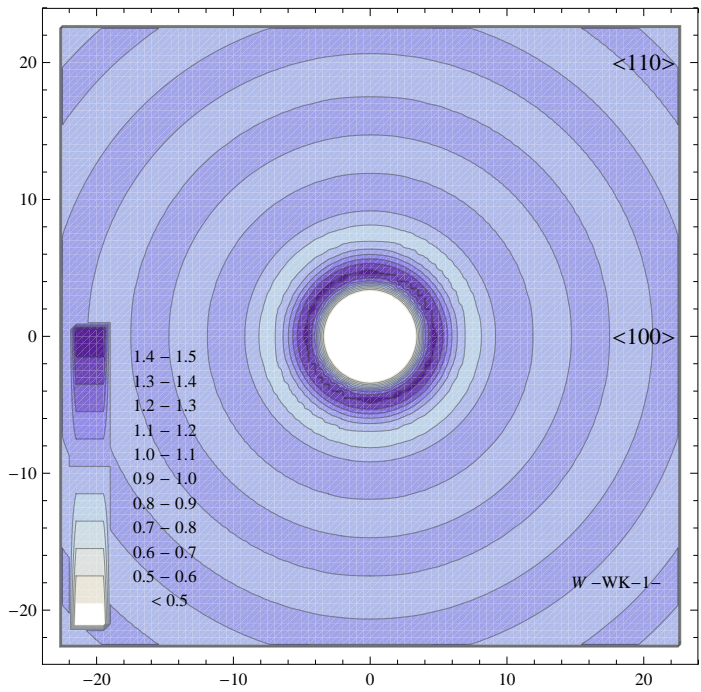

Fig. 3.c Limiting coverage g(s,$\theta_c$) contour plot for W

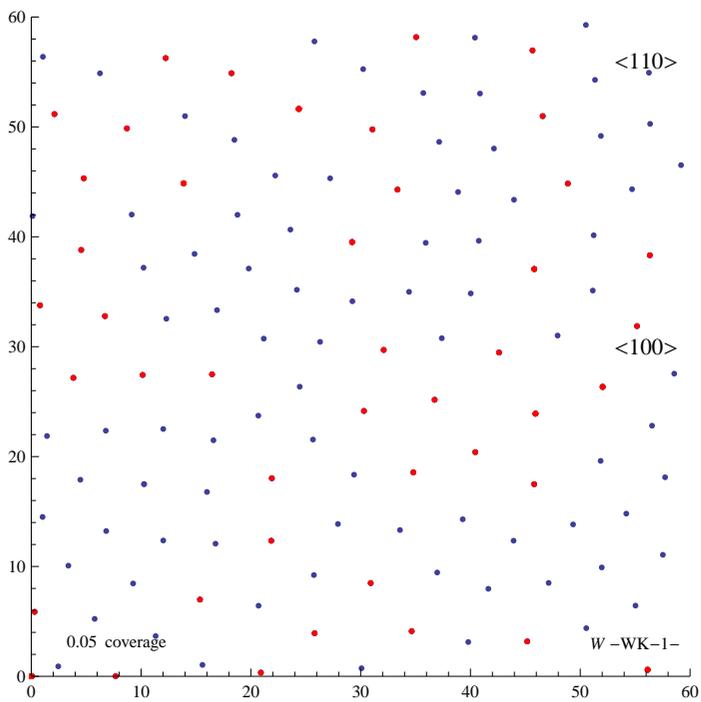

Fig. 3.d Limiting coverage adatom position sample for W

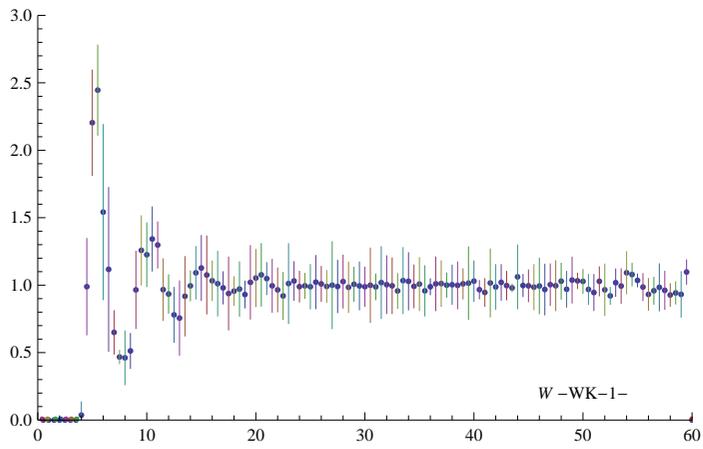

Fig. 3.e Pair correlation at limiting coverage from Monte Carlo simulation for W

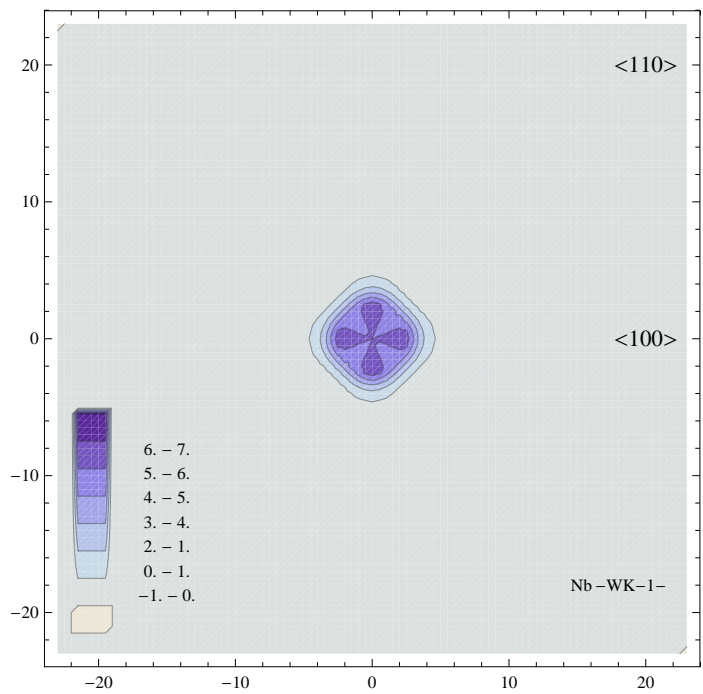

Fig. 4.a u(s) contour plot for Nb

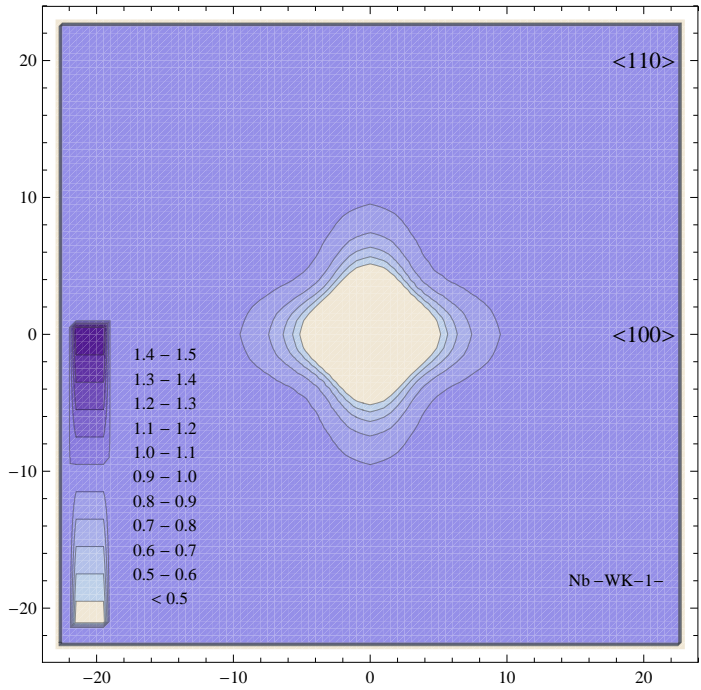

Fig. 4.b Zero coverage $g_0(s,0)$ contour plot for Nb

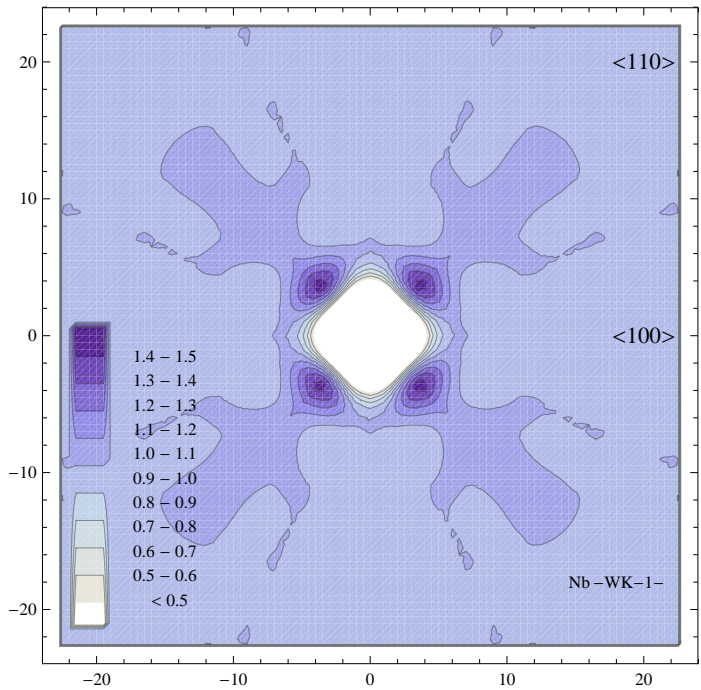

Fig. 4.c Limiting coverage $g(s,\theta_c)$ contour plot for Nb

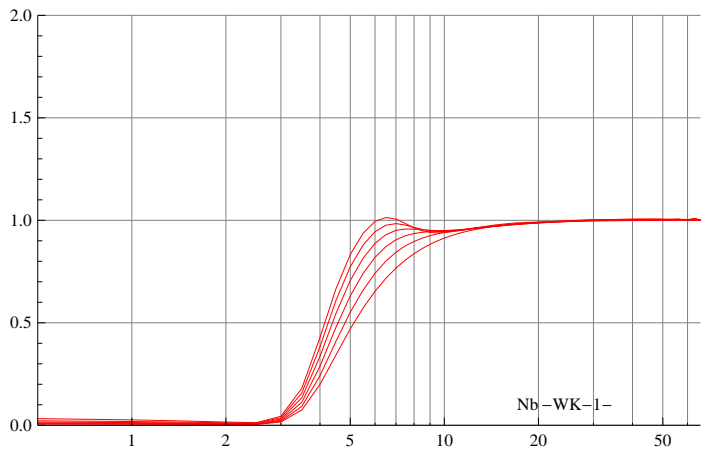

Fig. 4.d semi-logarithmic $g_{<100>}(s,\theta)$ plot for Nb, $\theta$ from 0 to 0.025.

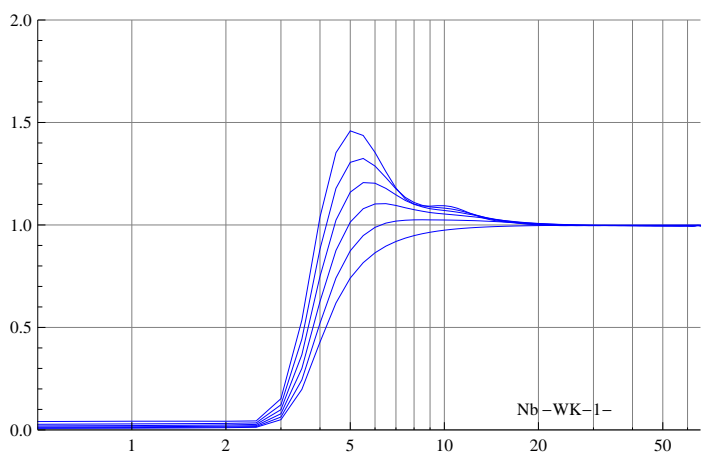

Fig. 4.e semi-logarithmic $g_{<110>}(s,\theta)$ plot for Nb, $\theta$ from 0 to 0.025.

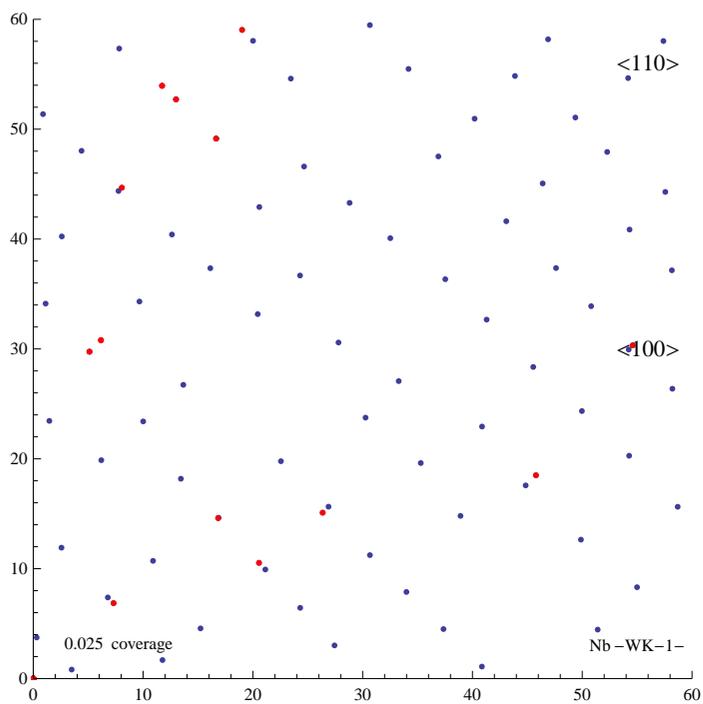

Fig. 4.f Limiting coverage adatom position sample for Nb

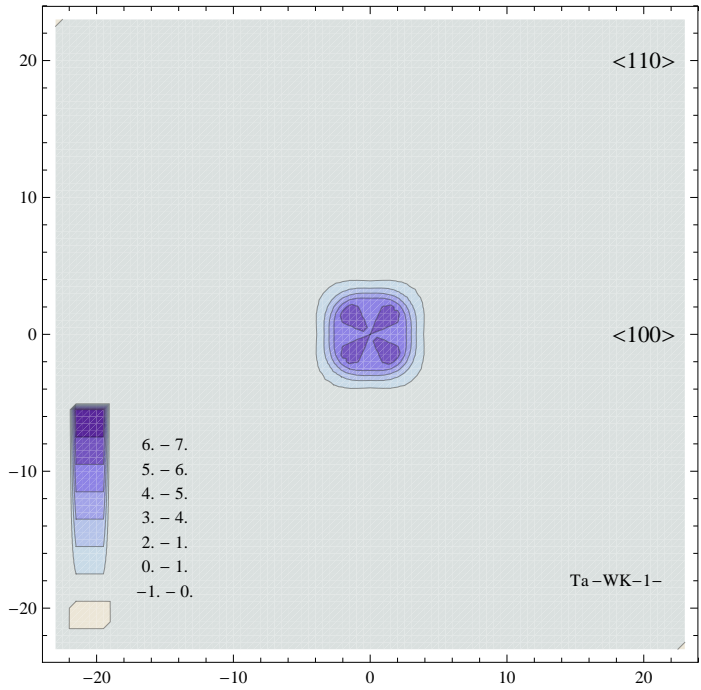

Fig. 5.a u(s) contour plot for Ta

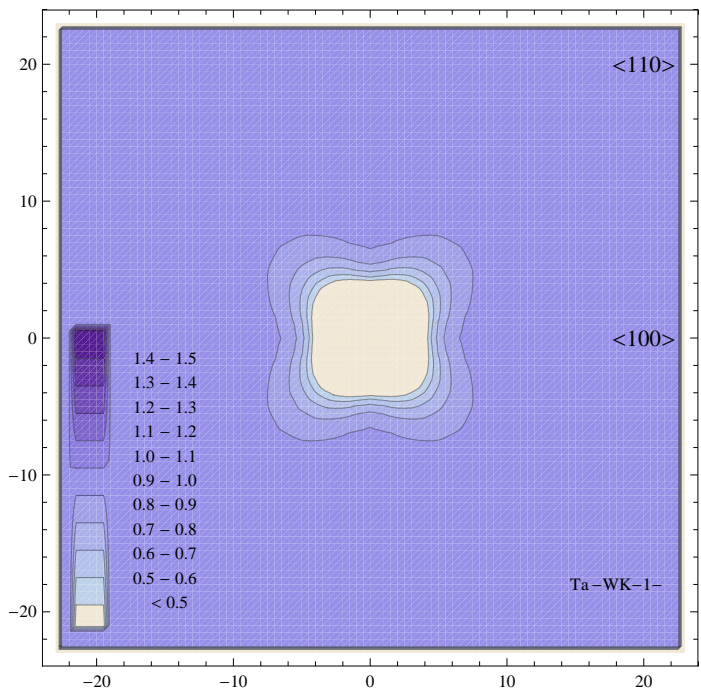

Fig. 5.b Zero coverage $g_0(s,0)$ contour plot for Ta

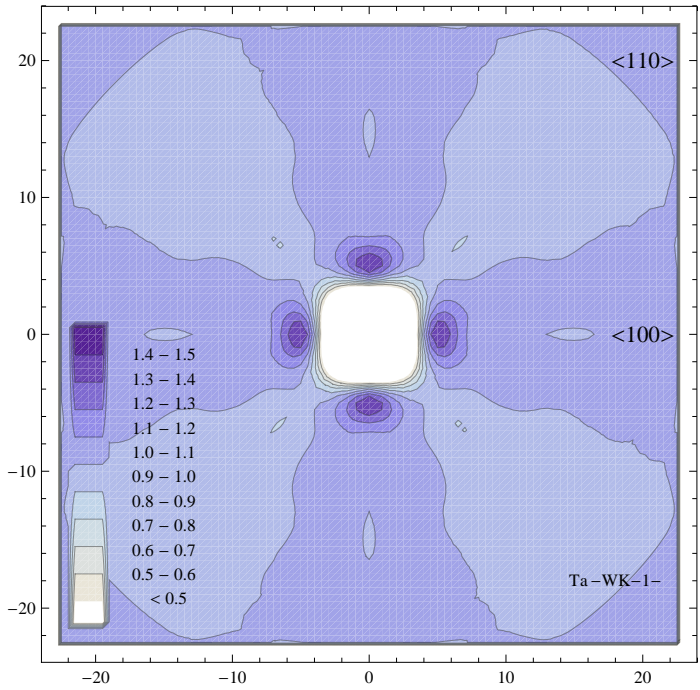

Fig. 5.c Limiting coverage g(s,θ_c) contour plot for Ta

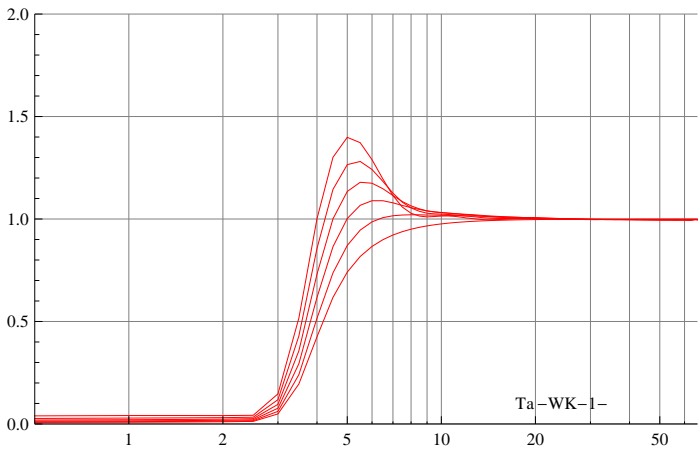

Fig. 5.d semi-logarithmic $g_{<100>}(s,\theta)$ plot for Ta, $\theta$ from 0 to 0.025.

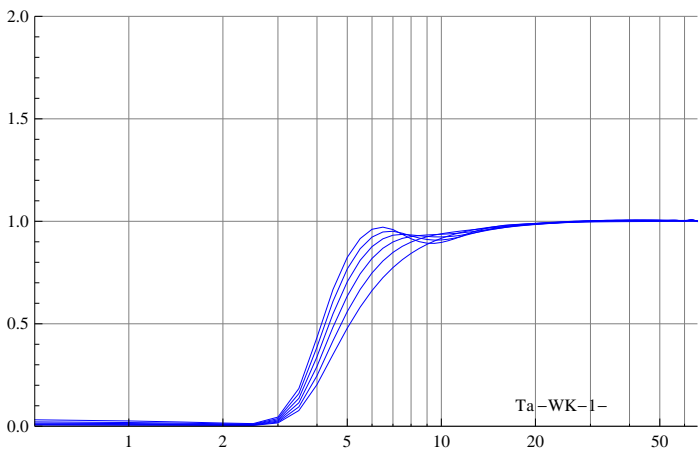

Fig. 5.e semi-logarithmic $g_{<110>}(s,\theta)$ plot for Ta, $\theta$ from 0 to 0.025.

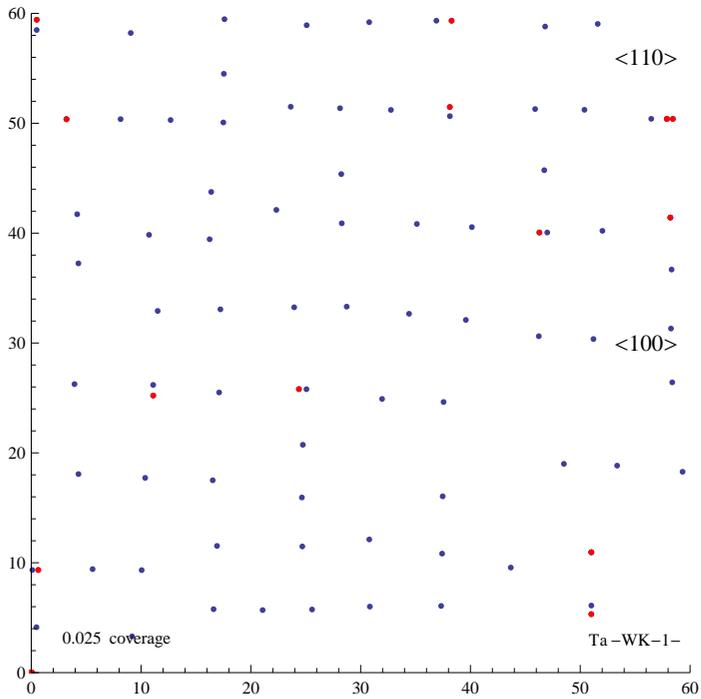

Fig. 5.f Limiting coverage adatom position sample for Ta

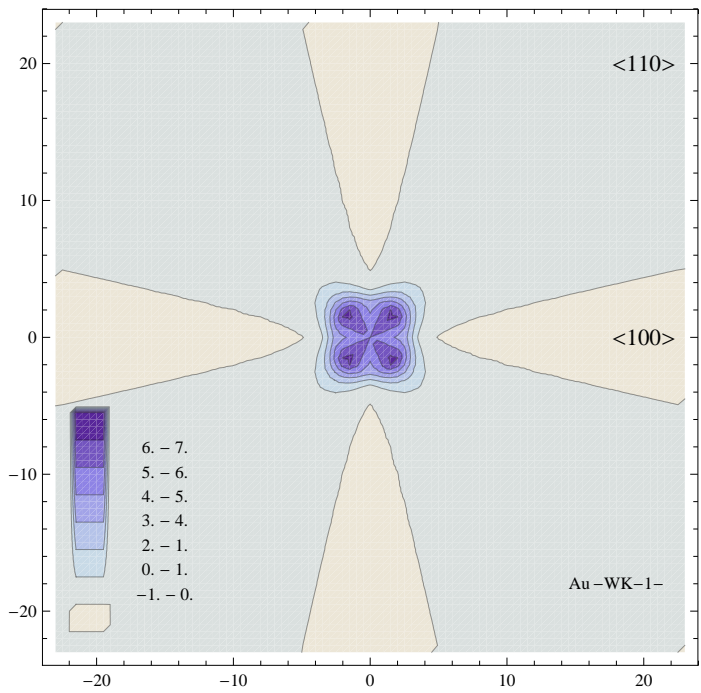

Fig. 6.a u(s) contour plot for <001> Au

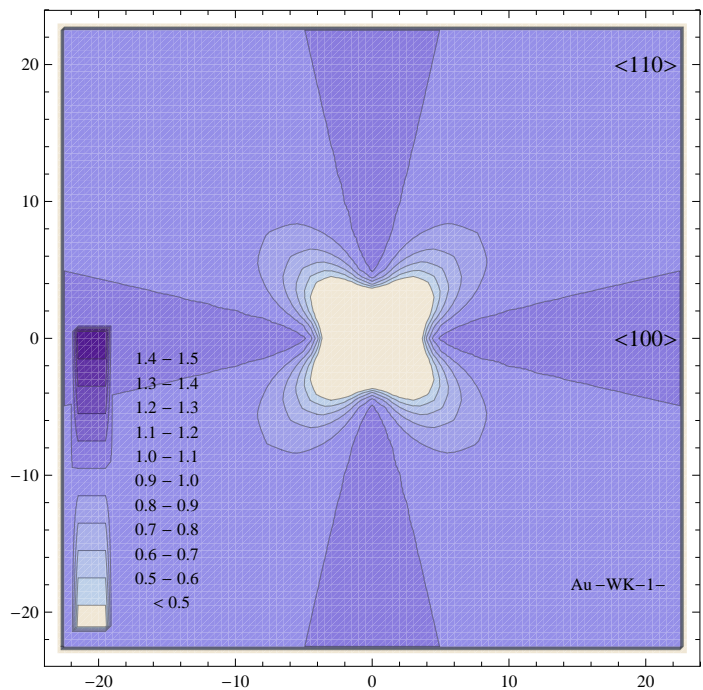

Fig. 6.b Zero coverage $g_0(s,0)$ contour plot for <001> Au

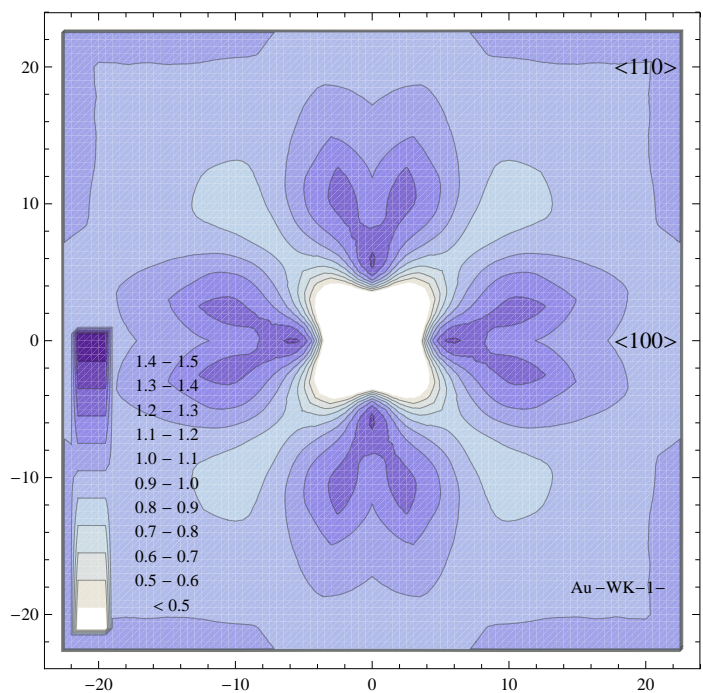

Fig. 6.c Limiting coverage $g(s,\theta_c)$ contour plot for <001> Au

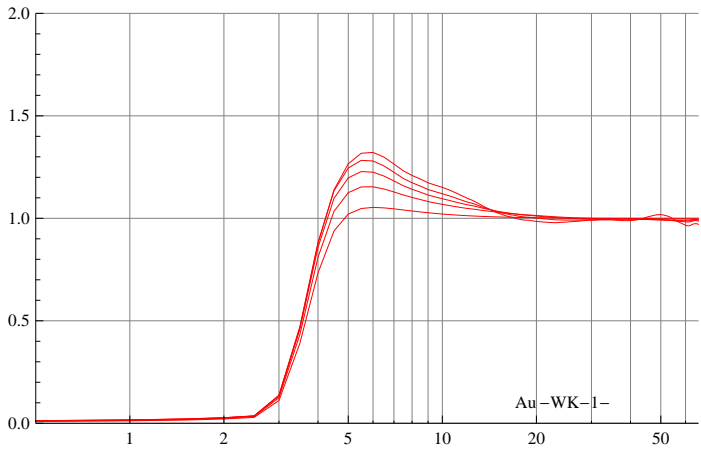

Fig. 6.d semi-logarithmic $g_{<100>}(s,\theta)$ plot for <001> Au, $\theta$ from 0 to 0.016.

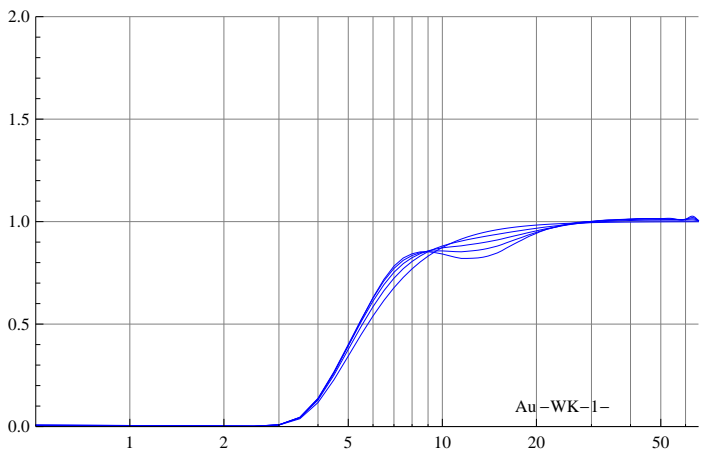

Fig. 6.e semi-logarithmic $g_{<110>}(s,\theta)$ plot for <001> Au, $\theta$ from 0 to 0.016.

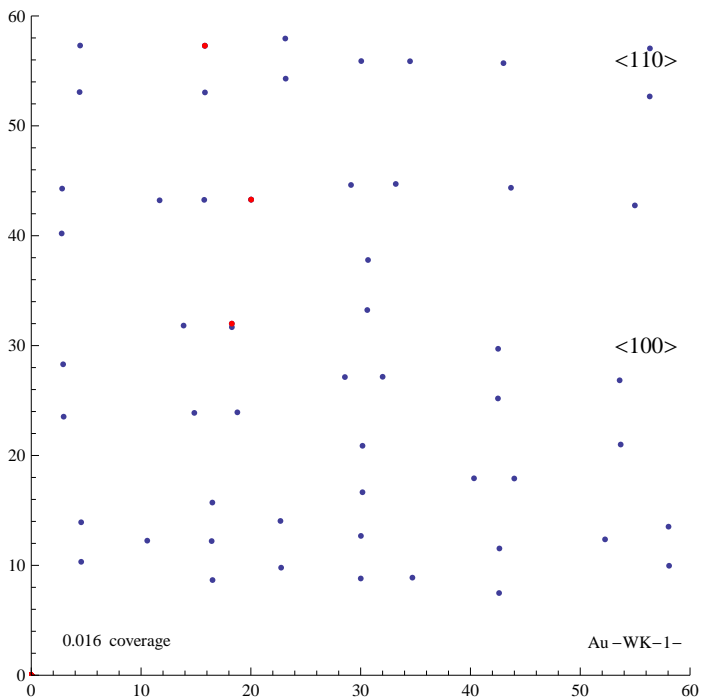

Fig. 6.f Limiting coverage adatom position sample for <001> Au

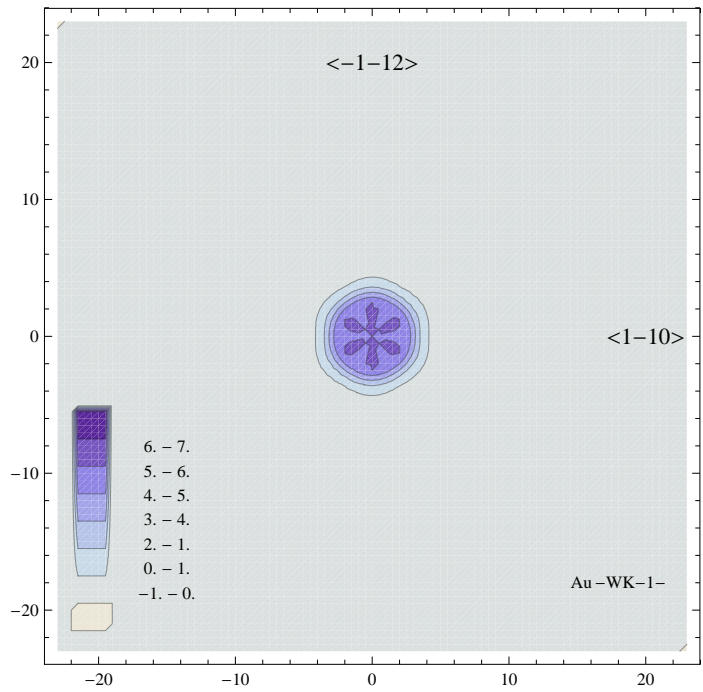

Fig. 7.a u(s) contour plot for <111> Au

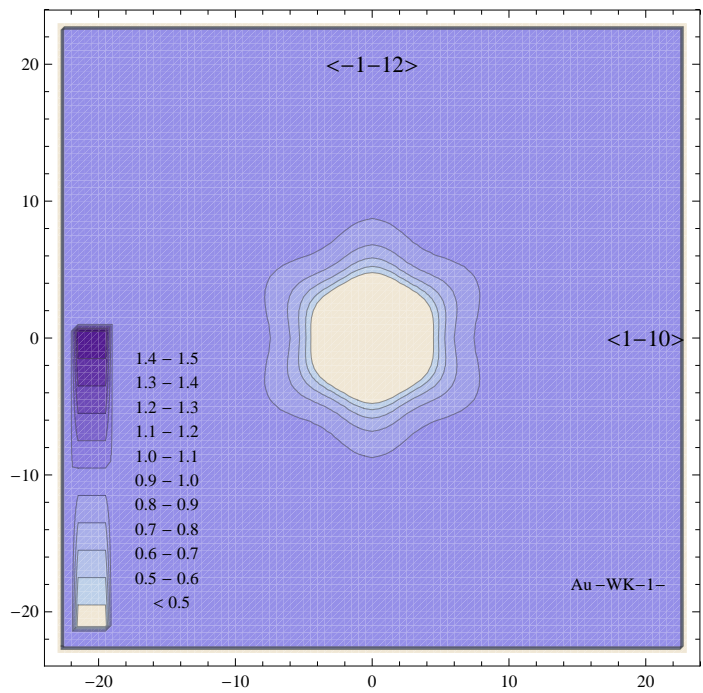

Fig. 7.b Zero coverage $g_0(s,0)$ contour plot for <111> Au

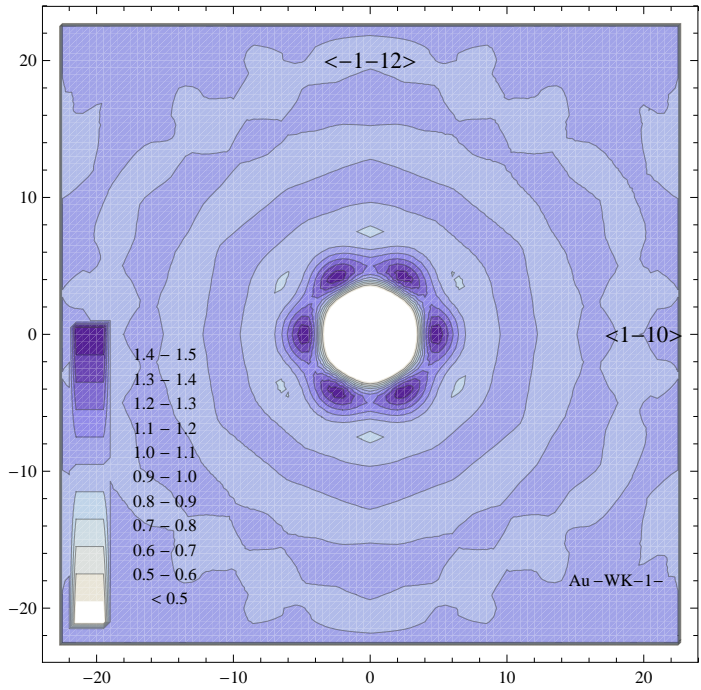

Fig. 7.c Limiting coverage $g(s,\theta_c)$ contour plot for <111> Au

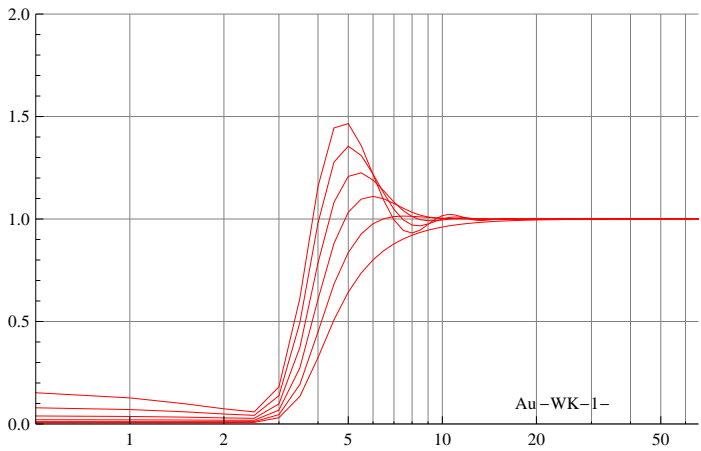

Fig. 7.d semi-logarithmic $g_{<1-10>}(s,\theta)$ plot for <111> Au, $\theta$ from 0 to 0.04.

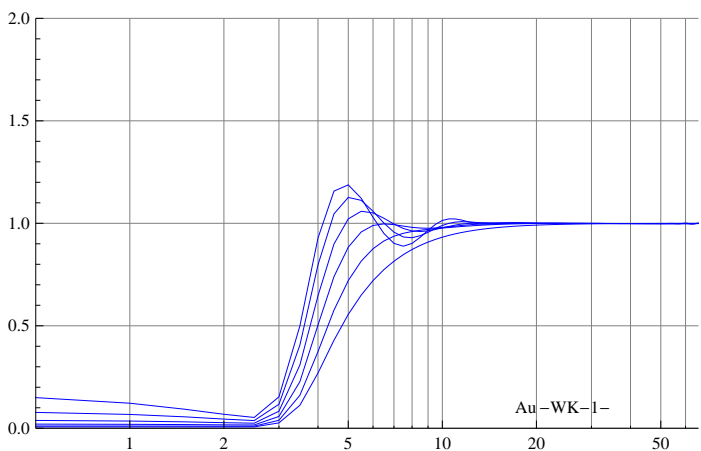

Fig. 7.e semi-logarithmic $g_{<1-12>}(s,\theta)$ plot for <111> Au, $\theta$ from 0 to 0.04.

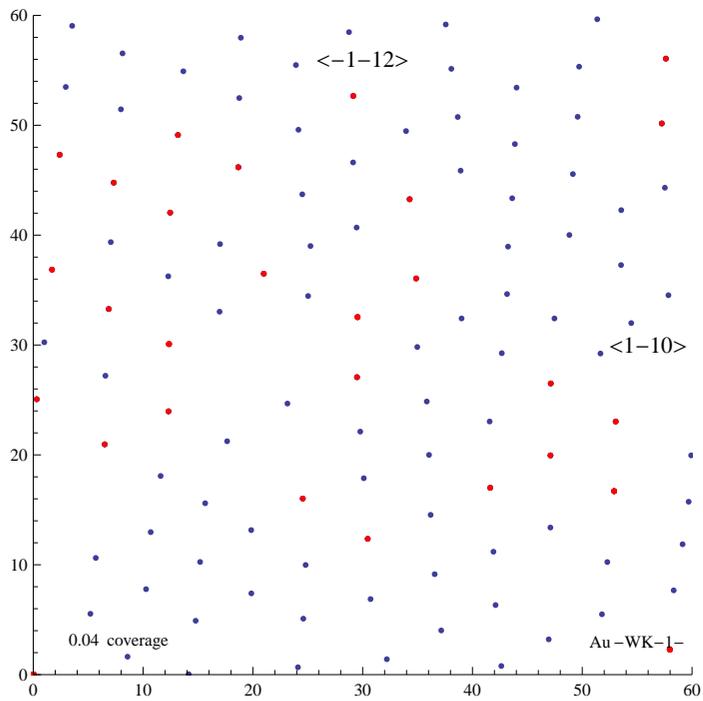

Fig. 7.f Limiting coverage adatom position sample for <111> Au